\documentclass{article}
\usepackage[left=1in,top=1in,right=1in,bottom=1in,head=.1in,nofoot]{geometry}

\usepackage{setspace,url,bm,amsmath} 
\usepackage{siunitx,booktabs,multirow, threeparttable}
\usepackage{titlesec} 
\usepackage{rotating}

\titlelabel{\thetitle.\quad} 
\titleformat*{\section}{\bf\large}

\usepackage{bbm, latexsym, xcolor, authblk}
\usepackage{amsthm, amsfonts}
\usepackage{comment}
\usepackage{appendix}

\theoremstyle{definition}

\theoremstyle{remark}
\newtheorem{remark}{Remark}[section]
\newtheorem{procedure}{Procedure}

\usepackage[authoryear, round]{natbib}
\bibpunct{(}{)}{;}{a}{,}{,} 
\usepackage{enumitem}

\usepackage{etoolbox} 
\apptocmd{\sloppy}{\hbadness 10000\relax}{}{}

\def\var{\textnormal{Var}}
\def\cov{\textnormal{Cov}}

\usepackage{algorithm}
\usepackage{algpseudocode}
\usepackage{graphicx, float, caption, subcaption}

\usepackage[listings]{tcolorbox}
\tcbuselibrary{theorems}

\newtcbtheorem[auto counter, number within=section]{boxthm}{Theorem}%
{colback=red!5,colframe=red!70,fonttitle=\bfseries, before skip=2em, after skip=1em}{th}

\newtcbtheorem[auto counter, number within=section]{boxdef}{Definition}%
{colback=cyan!5,colframe=cyan!50,fonttitle=\bfseries, before skip=2em, after skip=1em}{def}

\newtcbtheorem[auto counter, number within=section]{boxprop}{Proposition}%
{colback=gray!5,colframe=gray!50,fonttitle=\bfseries, before skip=2em, after skip=1em}{prop}

\usepackage{tikz}
\usetikzlibrary{positioning, arrows.meta, calc, shapes.geometric}

\title{\bf Variance Estimation Under Local Pivotal Methods}
\author[*]{Justin T. Greene}
\author[*]{Tirthankar Dasgupta}

\affil[*]{Department of Statistics, Rutgers University}

\begin{document}
\date{}
\doublespacing

\maketitle



\begin{abstract}

Spatially balanced sampling designs, such as the local pivotal methods (LPM), improve representation of auxiliary-variable space, but design-based variance estimation is challenging because second-order inclusion probabilities are generally intractable and may be zero or near zero. We propose two approaches for estimating the variance of the Horvitz–Thompson mean and the finite-population variance. The first estimates joint inclusion probabilities by Monte Carlo simulation and uses pairwise independence tests to modify the resulting variance estimators. The second, a model-assisted bootstrap (MABS), uses an interpolating Gaussian process to construct a synthetic finite population and estimates variance by resampling under the original LPM design. In simulations, the modified empirical approach performs well when responses are strongly associated with auxiliary variables, but its performance for the Horvitz–Thompson mean deteriorates as this association weakens. MABS exhibits more stable performance across noise levels and sampling settings, including under LPM2 and non-Gaussian noise. For finite-population variance estimation, however, no method shows a uniform advantage. These results illustrate the complementary strengths of the proposed approaches for inference under spatially balanced sampling.

\end{abstract}


\section{Introduction} \label{sec:intro}
Simple random sampling (SRS) is the standard design for unbiased estimation of a finite population mean. When no auxiliary information is available, simple random sampling is difficult to improve. However, often times each unit in a population has features/additional measurements that are correlated with the response of interest; we call these \textit{auxiliary variables}. A simple random sample may fail to represent the full range of features, leaving regions of auxiliary space overrepresented while others are ignored. Despite remaining unbiased, this imbalance inflates the variance of the mean estimator. As a concrete example, suppose we want to estimate the average grade $Y$ on a writing assignment across $N$ students. For each essay we observe two text features: (1) average sentence length and (2) lexical diversity, measured as the ratio of unique words to total words. Essays with similar text features will likely receive similar grades, thus a sample that neglects regions of the auxiliary space will produce a less precise estimate than a sample that spreads evenly across the text features. 

A natural first step in using auxiliary variables is to define strata with respect to similarity in the auxiliary distributions among the units in the population. Then, we can simply use stratified random sampling in which a random sample is taken from each strata \citep{neyman1934}. As the dimension of the auxiliary space increases, explicitly defining strata becomes more challenging. Space-filling designs, such as Latin hypercube designs \citep{lin2022, minasny2006}, are commonly used to achieve stratification by evenly spreading units across the auxiliary space. However, these designs are used to select points from a continuous design space, but typically not for selecting a subset from a finite-population of points within auxiliary space. 

Sampling methods that preserve first-order inclusion probabilities while taking auxiliary data into account include the cube, local pivotal, and local cube methods \citep{deville2004, grafstrom2012, grafstrom2013localcube}. The local pivotal methods provide samples that are \textit{well spread} over the space of auxiliary variables; they avoid selecting neighboring units in the same sample. Since well spread samples in the auxiliary space are also approximately balanced on the auxiliary variables \citep{grafstrom2013}, spatial balance subsumes the balancing goals of the cube and local cube methods without additional computational complexity. 

The local pivotal methods respect fixed first-order inclusion probabilities $\pi_i$ for $i=1,\dots,N$. Since these methods do not require each $\pi_i$ to be equal, an applicable unbiased estimator of the population mean is the Horvitz-Thompson mean \citep{horvitz1952}. Unbiased estimators of the variance of the Horvitz-Thompson mean are defined by \citet{horvitz1952} as well as \citet{sen1953, yates1953}; with the latter being preferred because it guarantees non-negativity. Similarly, population variance estimates are presented by \citet{liu1974} and \citet{chaudhuri1978}, with Chaudhuri's estimator ensuring non-negativity. 

The key issue with these unbiased variance estimators is the requirement for the second-order inclusion probabilities $\pi_{ij}$ to be known and strictly positive. With the local pivotal methods, the second-order inclusion probabilities are intractable. Even if they had a closed-form equation, as a consequence of inducing negative dependence between nearby pairs, many of these probabilities would be zero or extremely small. When $\pi_{ij} = 0$, units $i$ and $j$ can never appear jointly in the realized sample. Consequently, although their pairwise difference contributes to the variance quantity being estimated, that contrast can never appear in the corresponding sample-based estimator, producing downward bias whenever the corresponding contrast is nonzero. 

Existing alternatives include empirically estimating the second-order inclusion probabilities using $M$ samples taken with the desired sampling design \citep{fattorini2006} or using variance estimators that forgo the use of second-order inclusion probabilities by instead using variance within local neighborhoods \citep{wolter2007, grafstrom2014, grafstrom2018, olofsson2023,stevens2003}. Our numerical explorations show that these alternatives can lead to overestimation of the variance, making it difficult for the variance reduction provided by the local pivotal methods to be realized when making inferences.

In this paper, we develop two approaches to variance estimation under the local pivotal methods that aim to address these challenges while retaining the advantages of spatially balanced sampling. Our first approach builds on the empirical estimation of second-order inclusion probabilities and introduces a modification that accounts for the negative dependence induced by the sampling design. Specifically, we test whether each pair of sampling indicators behaves independently; when independence cannot be rejected, we set the empirical estimate of the joint inclusion probability equal to the product of the corresponding first order inclusion probabilities, otherwise we retain the empirical estimate. This modification is motivated by the observation that empirical estimates can exhibit substantial variability for pairs that are effectively independent, resulting in artificial contributions to the estimated variance. We use this procedure to construct modified versions of the Sen–Yates–Grundy estimator for the variance of the Horvitz–Thompson mean and the Chaudhuri estimator for the finite population variance.

Our second approach, which we call the model-assisted bootstrap (MABS), avoids the direct estimation of second-order inclusion probabilities altogether. We fit a Gaussian process model to the observed responses and auxiliary variables of the sampled units, using the fitted model to construct a synthetic finite population that retains the observed responses at sampled units while accounting for predictive uncertainty at nonsampled units. We then repeatedly draw bootstrap samples from this synthetic population using the original local pivotal method sampling design and estimate the variance of the Horvitz–Thompson mean from the resulting bootstrap replicates. We also use the synthetic population to estimate the finite population variance. 

An important contribution of this work is a systematic comparison of the proposed methods with existing variance estimators, including neighborhood-based estimators and estimators based on empirical second-order inclusion probabilities. Our simulation results reveal complementary features of the two proposed approaches. The modified empirical inclusion-probability approach performs well when the relationship between the auxiliary variables and the response is strong, but its performance deteriorates as the signal-to-noise ratio decreases. In contrast, the MABS approach exhibits greater robustness to increasing noise, maintaining relatively low bias and mean squared error across the range of signal-to-noise ratios considered. We further investigate the robustness of MABS to changes in the sampling design and the response-generating process, including a second variant of the local pivotal method (LPM2), non-Gaussian noise, and an alternative prediction model.

The paper is organized as follows. Section \ref{sec:background} introduces the notation and background on local pivotal methods, reviews inference for the finite population mean and variance, and describes existing alternatives when second-order inclusion probabilities are unknown. Section \ref{sec:method1} develops the proposed variance estimators based on empirical estimation of second-order inclusion probabilities, including the modification that incorporates a hypothesis test for independence between pairs. Section \ref{sec:method2} introduces the model-assisted bootstrap approach and describes the construction of the interpolating Gaussian process predictor. Section \ref{sec:validating_sim_study} presents a simulation study comparing the proposed estimators with existing alternatives under different signal-to-noise ratios, sampling designs, and prediction models. Finally, Section \ref{sec:discussion} provides a discussion of the findings and their implications.


\section{Notation, Background, and Definitions} \label{sec:background}


\subsection{Local Pivotal Methods} \label{subsec:lpms}
Let $\mathcal{U} = \{1,\dots,N\}$ denote a finite population of $N$ units from which we take a random sample of $n$ units  $\mathcal{S} \subset \mathcal{U}$ to estimate the population mean. Let $S_i$ be an indicator random variable, with $S_i = 1$ if unit $i$ is sampled (i.e. $i \in \mathcal{S}$) and $S_i = 0$ otherwise ($i \notin \mathcal{S}$). The joint probability distribution of $(S_1,\dots, S_N)$, denoted by $p(\cdot)$, is governed by the \emph{sampling design}. Several examples of sampling designs can be found in \citet{hansen1953, kish1965, cochran1977}. The probability that unit $i$ is included in the sample is denoted as the first-order inclusion probability $\pi_i = \Pr(S_i = 1)$ and the probability units $i$ and $j$ are included in the same sample is the joint second-order inclusion probability $\pi_{ij} = \Pr(S_i = S_j = 1)$. For example, in simple random sampling without replacement, $\pi_i = n/N$ for all $i=1, \ldots, N$ and $\pi_{ij} = n(n-1)/N(N-1)$ for $i \ne j, i=1, \ldots, N, j=1, \ldots, N.$ 

Here we assume that a vector $\mathbf{X}_i$ of auxiliary variables is available for each unit $i=1, \ldots, N$ prior to the sampling. The goal is to select a sample that is well spread over the auxiliary space. The sampling designs we focus on are the local pivotal methods \citep{grafstrom2012}. These methods are algorithms that take the vector of first-order inclusion probabilities $(\pi_1,\dots,\pi_N)$ with $\sum_{i=1}^N \pi_i = n$ as input and return a sample $\mathcal{S} \subset \{1,\dots,N\}$ of exactly $n$ units, where each unit $i$ is included with a probability of $\pi_i$ \citep{grafstrom2012}. The resultant sample is expected to represent the auxiliary space well. Algorithm \ref{alg:LPM1} outlines the pseudocode for local pivotal method 1 (LPM1). The key intuition of the local pivotal methods are that they initialize decision markers $\delta_i = \pi_i$ for $i=1,\dots,N$ and then repeatedly update these decision markers until $n$ of them have $\delta_i = 1$, with $\mathcal{S} = \{i : \delta_i = 1\}$ making up the sampled units. Proximity between two points $(i,j)$ is measured by a distance (typically Euclidean) between auxiliary vectors $\mathbf{X}_i$ and $\mathbf{X}_j$. At each step, two nearby decision markers $(\delta_i, \delta_j)$ are selected. If $\delta_i + \delta_j < 1$, then we choose to \textit{exclude} one of the units, while if $\delta_i +\delta_j \geq 1$, we \textit{include} one of the units.

The updating rules in Algorithm \ref{alg:LPM1} (lines 6-9), are directly from the pivotal method defined in \citet{deville1998}. \citeauthor{deville1998} used the updating rules on a randomly chosen pair $(i,j)$, while \citeauthor{grafstrom2012} used it on nearby units at each step. \citet{grafstrom2012} also introduced the local pivotal method 2 (LPM2), which drops the need for unit $j$'s nearest neighbor to be unit $i$ (line 5 in Algorithm \ref{alg:LPM1}). By dropping this mutual nearest neighbor condition there are no longer iterations where unit $i$ is selected, the nearest neighbor search is completed to identify $j$, and no updating is conducted. For this reason, LPM2 can be used if faster computation is needed. 

\begin{algorithm}
\caption{Local Pivotal Method 1 (LPM1)}
\label{alg:LPM1}
\begin{algorithmic}[1]
  \Require Population size $N$, sample size $n$, auxiliary vars $\{\bm{X}_i\}_{i=1}^N$, inclusion probs $\{\pi_i\}_{i=1}^N$ with $\sum_i \pi_i = n$, distance measure $d(\bm{X}_i , \bm{X}_j)$
  \State \textbf{Initialize:} $\delta_i = \pi_i$ for all $i = 1,\dots,N$;\;
         $\mathcal{A} = \{1,\dots,N\}$
  \While{$|\mathcal{A}| > 1$}
    \State Pick $i$ uniformly at random from $\mathcal{A}$
    \State Let $j$ be the nearest neighbor of $i$ in $\mathcal{A} \setminus\{i\}$ under $d(\cdot, \cdot)$ (break ties uniformly at random)
    \If{$i$ is also the nearest neighbor of $j$ in $\mathcal{A} \setminus\{j\}$ under $d(\cdot, \cdot)$}
      \If{$\delta_i + \delta_j < 1$}
        \State
        $(\delta_i, \delta_j) = \begin{cases}
          (0,\; \delta_i + \delta_j) & \text{with probability } \dfrac{\delta_j}{\delta_i + \delta_j} \\[6pt]
          (\delta_i + \delta_j,\; 0) & \text{with probability } \dfrac{\delta_i}{\delta_i + \delta_j}
        \end{cases}$
      \ElsIf{$\delta_i + \delta_j \geq 1$}
        \State
        $(\delta_i, \delta_j) = \begin{cases}
          (1,\; \delta_i + \delta_j - 1) & \text{with probability } \dfrac{1 - \delta_j}{2 - \delta_i - \delta_j} \\[6pt]
          (\delta_i + \delta_j - 1,\; 1) & \text{with probability } \dfrac{1 - \delta_i}{2 - \delta_i - \delta_j}
        \end{cases}$
      \EndIf
      \State $\mathcal{A} \leftarrow \mathcal{A} \setminus \{k \in \{i,j\} : \delta_k \in \{0,1\}\}$
    \EndIf
  \EndWhile
  \State \Return $\mathcal{S} = \{k : \delta_k = 1\}$
\end{algorithmic}
\end{algorithm}

\begin{remark}\label{remarklpm}
    When using a scale-sensitive distance measure, the auxiliary variables should be standardized or otherwise placed on comparable scales so that no variable dominates the distance solely due to its scale.
\end{remark}


\subsection{Inference for finite population mean} \label{subsec:finite_pop_estimand}
Let $Y_i \in \mathbb{R}$ denote the fixed response for unit $i=1, \ldots, N$. Our estimand of interest is the finite population mean:
\begin{equation}
    \overline{Y} = \dfrac{1}{N}\sum_{i=1}^N Y_i. \label{eq:pop_mean}
\end{equation}

In general, to estimate the population mean $\overline{Y}$ under sampling designs with fixed first-order inclusion probabilities $\pi_i = \Pr(S_i = 1)$ for $i=1,\dots,N$, we rely on the Horvitz-Thompson (HT) estimator of population mean \citep{horvitz1952}
\begin{equation}
    \overline{y} = \dfrac{1}{N}\sum_{i=1}^N S_i \dfrac{Y_i}{\pi_i}. \label{eq:ht_mean}
\end{equation}
The HT estimator is design-unbiased, meaning $\mathbb{E}_p[\overline{y}] = \overline{Y}$, where $\mathbb{E}_p[\cdot]$ is the expectation with respect to sampling design $p(\cdot)$. Notably, when taking simple random samples, the first-order inclusion probabilities are all equal $\pi_i = n/N$ and the HT mean simplifies to the unweighted sample mean, which we will denote as $\overline{y}_{\mathrm{SRS}} = \left(\sum_{i = 1}^N S_i Y_i \right)/n$. 

The variance of the HT mean is \citep{horvitz1952}:
\begin{equation}
    \var_p(\overline{y}) = \frac{1}{N^2} \bigg[ \sum_{i=1}^N (1 - \pi_i) \frac{Y_i^2}{\pi_i} + \sum_{i=1}^N \sum_{\substack{j=1 \\ j \ne i}}^N (\pi_{ij} - \pi_i \pi_j) \frac{Y_i}{\pi_i}\frac{Y_j}{\pi_j} \bigg]. \label{eq:true_ht_mean_var}
\end{equation}
While \citet{horvitz1952} proposed an unbiased estimator for (\ref{eq:true_ht_mean_var}), it can result in negative variance estimates. To ensure stable, non-negative variance estimates for fixed-size designs, like the local pivotal methods, we instead use the Sen-Yates-Grundy (SYG) variance estimator \citep{sen1953, yates1953}:
\begin{equation}
    \widehat{\var}_{\mathrm{SYG}}(\overline{y}) = \dfrac{1}{N^2}\sum_{i=1}^{N-1}\sum_{j = i +1}^{N} S_i S_j \dfrac{\pi_i \pi_j - \pi_{ij}}{\pi_{ij}} \bigg( \dfrac{Y_i}{\pi_i} - \dfrac{Y_j}{\pi_j} \bigg)^2. \label{eq:syg_var}
\end{equation}
Provided $\pi_{ij} > 0$ for all pairs, this is an unbiased estimator of (\ref{eq:true_ht_mean_var}). The SYG variance estimator is guaranteed to be non-negative when $\pi_{ij} \leq \pi_i \pi_j$ for all pairs, which is the case for the local pivotal methods. Under SRS, the SYG variance simplifies to the standard finite-population corrected estimator for variance of the sample mean; $\var(\overline{y}_{\mathrm{SRS}}) = (1 - n/N)\ (S^2/n)$ where $S^2$ is the finite population variance with divisor $N-1$ as defined in (\ref{eq:pop_var}). 


\subsection{Inference for finite population variance}\label{subsec:pop_var}

Another population summary of interest is the true spread of the outcomes $Y_1, \ldots, Y_N$, or the finite population variance:
\begin{equation}
    S^2 = \dfrac{1}{N-1}\sum_{i=1}^N (Y_i - \overline{Y})^2. \label{eq:pop_var}
\end{equation}
The above equation can be expanded into the following two quadratic forms \citep{cochran1977}: 
\begin{equation}
   S^2 = \dfrac{1}{N-1}\bigg[\sum_{i=1}^N Y_i^2 - \dfrac{1}{N}\bigg( \sum_{i=1}^N Y_i \bigg)^2  \bigg] = \dfrac{1}{N}\sum_{i=1}^N Y_i^2 - \dfrac{1}{N (N-1)}\sum_{i=1}^N \sum_{\substack{j=1 \\ j \ne i}}^N Y_i Y_j. \label{eq:quad_pop_var}
\end{equation}
Under SRS, since $\overline{y} = \overline{y}_{\mathrm{SRS}}$, the unweighted sample variance is unbiased
\begin{equation}
    s_{\mathrm{naive}}^2 = \dfrac{1}{n-1}\sum_{i=1}^N S_i (Y_i - \overline{y})^2. \label{eq:naive_pop_var}
\end{equation}
We call this the naive estimator because it is biased for sampling designs like the local pivotal methods. Another estimator of population variance is obtained by replacing the unknown population totals $\sum_{i=1}^N Y_i$ and $\sum_{i=1}^N Y_i^2$ in the first quadratic form of (\ref{eq:quad_pop_var}) with their Horvitz-Thompson estimators
\begin{equation}
    s_{\mathrm{HT}}^2 = \dfrac{1}{N-1}\bigg[ \sum_{i=1}^N S_i \dfrac{Y_i^2}{\pi_i} - \dfrac{1}{N}\bigg( \sum_{i=1}^N S_i \dfrac{Y_i}{\pi_i} \bigg)^2 \bigg]. \label{eq:ht_pop_var}
\end{equation}

Neither of the estimators \eqref{eq:naive_pop_var} or \eqref{eq:ht_pop_var} require knowledge of the second-order inclusion probabilities. However, both estimators are biased. We will evaluate their performance in the simulation section later.

To derive an unbiased estimator of $S^2$, \citet{liu1974} applied the Horvitz-Thompson principle of using weighted response variables to the second quadratic form of the population variance in (\ref{eq:quad_pop_var}). The Liu estimator is unbiased, but can result in negative estimates. \citet{chaudhuri1978} proposed a non-negative, unbiased estimator (when all $\pi_{ij} > 0$) of the population variance, which we will call the ``Chaud'' estimator
\begin{equation}
    s_{\mathrm{Chaud}}^2 = \dfrac{1}{N(N-1)}\sum_{i=1}^{N-1}\sum_{j = i+1}^N S_i S_j \dfrac{(Y_i - Y_j)^2}{\pi_{ij}} \label{eq:chaud_pop_var}.
\end{equation}
The above equation corresponds to $t_3$ in \citet{chaudhuri1978} rescaled by $N / (N-1)$ to accommodate for \citeauthor{chaudhuri1978} using $1/N$ in their equation for $S^2$. 

\begin{remark}\label{remark_pi_ij_zero}
    Both the SYG (\ref{eq:syg_var}) estimator for the variance of HT mean and Chaudhuri (\ref{eq:chaud_pop_var}) estimator of population variance require $\pi_{ij} > 0$ for all $\binom{N}{2}$ pairs in the population to remain unbiased. When $\pi_{ij} = 0$, units $i$ and $j$ can never be selected jointly, so the corresponding pairwise contrast can never contribute to the estimator in any realized sample. Thus, if pairs with $\pi_{ij} = 0$ have nonzero differences, their contributions to the target variance cannot be recovered, leading to downward bias. For the Chaudhuri, the missing contribution to $S^2$ is proportional to $(Y_i - Y_j)^2$; for the SYG, it is proportional to $\pi_i\pi_j(Y_i/\pi_i - Y_j / \pi_j)^2$. We will see in Section \ref{ss:motivating_sim_study} that the local pivotal method can result in $\pi_{ij} = 0$.
\end{remark}


\subsection{Alternatives when second-order inclusion probabilities are unknown} \label{subsec:alt_variance_ests}
Without knowledge of the $\pi_{ij}$, there are two main alternatives past work has employed: (a) estimating variance of the HT mean estimator within local neighborhoods of sampled units and (b) estimating the second-order inclusion probabilities. We discuss these two approaches in the following subsections. 


\subsubsection{Neighborhood-based methods}
The local mean variance estimator \citep{grafstrom2014, grafstrom2018} is defined as
\begin{equation}
    \widehat{\text{Var}}_{n^{\prime}-\text{NN}}(\overline{y}) = \frac{1}{N^2}  \sum_{i=1}^N S_i \cdot \frac{n_i}{n_i - 1}\bigg( \frac{Y_i}{\pi_i} - \frac{1}{n_i} \sum_{j \in \mathcal{N}_i} \frac{Y_j}{\pi_j} \bigg)^2  \label{eq:local_mean}
\end{equation}
where $\mathcal{N}_i$ is the neighborhood of unit $i$, including the indices for unit $i$ and its $n^{\prime} - 1$ nearest sampled neighbors, and $n_i = |\mathcal{N}_i|$ is the neighborhood size. If multiple units are tied at the distance of the $(n^{\prime} - 1)^{th}$ nearest neighbor, all tied units are included in $\mathcal{N}_i$. Thus, $n_i \geq n^{\prime}$, with $n_i = n^{\prime}$ in the absence of ties. This estimator is recommended by \citet{grafstrom2014, grafstrom2018, olofsson2023}, with \citet{olofsson2023} noting that thorough investigation is needed to find the optimal neighborhood size $n^{\prime}$ for a given finite population and sample size. For the lowest bias, \citet{grafstrom2018} recommends using (\ref{eq:local_mean}) when $n^{\prime} = 2$, which when $n_i = 2$ for all sampled units simplifies to what we will call the nearest neighbor (NN) estimate
\begin{equation}
     \widehat{\text{Var}}_{\text{NN}}(\overline{y}) = \frac{1}{N^2} \cdot  \frac{1}{2}\sum_{i=1}^N  S_i \bigg( \frac{Y_i}{\pi_i} - \frac{Y_{j_i}}{\pi_{j_i}}\bigg)^2  \label{eq:near_neigh}
\end{equation}
where $j_i$ is the index of unit $i$'s nearest neighbor in the sample. Lastly, \citet{stevens2003} define a neighborhood based variance estimate that includes weights,
\begin{equation}
    \widehat{\text{Var}}_{\text{NBH}}(\overline{y}) = \frac{1}{N^2}\sum_{i=1}^N \sum_{j \in D(i)} S_i w_{ij}\bigg( \frac{y_j}{\pi_j} - \sum_{k \in D(i)} w_{ik} \frac{y_k}{\pi_k}\bigg)^2 \label{eq:NBH_var}
\end{equation}
where $w_{ij}$ are the optimized weights and $D(i)$ is the symmetrized neighborhood of unit $i$. Readers are referred to \cite{stevens2003} for details on weight calculation.

These alternatives [(\ref{eq:local_mean}), (\ref{eq:near_neigh}), (\ref{eq:NBH_var})] for estimating $\var(\overline{y})$ and their pitfalls will be explored in Section \ref{ss:motivating_sim_study}.  The neighborhood-based methods do not apply to the problem of estimation of the sampling variance $S^2$. In the absence of knowledge of second-order inclusion probabilities, available options include biased estimators (\ref{eq:naive_pop_var}) and (\ref{eq:ht_pop_var}).


\subsubsection{Estimation of second-order inclusion probabilities for LPM} \label{sec:estimating_pi_ij_previous_attempts}
Previous work has been done to estimate the second-order inclusion probabilities using the first-order probabilities \citep{Hartley1962, Hajek1964, tille1996, Brewer2003}. For example, \citet{Hajek1964} approximated $\pi_{ij}$ with 
\begin{equation}
    \widehat\pi_{ij} = \pi_i \pi_j \frac{1 - (1 - \pi_i)(1-\pi_j)}{\sum_{i=1}^N \pi_i (1 - \pi_i)}. \label{eq:pi_ij_high_entropy} 
\end{equation}
However, these estimates that utilize the first-order inclusion probabilities rely on the sampling design having high entropy. Since local pivotal methods make selecting nearby pairs in the same sample less likely, many of the $\binom{N}{n}$ possible samples are highly unlikely to be observed. Thus, the local pivotal methods do not have high entropy, making these second-order inclusion probability estimates unlikely to yield results. 

\citet{fattorini2006} explored the use of estimating the second-order inclusion probabilities empirically over $M$ independent Monte Carlo replications of taking a sample of size $n$ from $N$ units with the specific sampling design as
\begin{equation}
    \widehat{\pi}^{\text{F}}_{ij} = \dfrac{C_{ij} + 1}{M + 1} \label{eq:emp_pi_ij_unb}
\end{equation}
where $C_{ij}$ is the count of times units $i$ and $j$ co-occurred in a sample throughout the $M$ replications. The estimated second order inclusion probabilities can be substituted in \eqref{eq:syg_var} and \eqref{eq:chaud_pop_var} to respectively estimate the variances of the HT estimator and population variance.

There are two issues associated with LPM sampling that demand some investigation (and possible modifcation) of the above procedure before its direct application. First, since the LPMs only induce negative dependence between units, we know there is an upper bound $\pi_{ij} \leq \pi_i \pi_j$. Additionally, the second-order inclusion probabilities are the intersection of two events $S_i = 1$ and $S_j = 1$. Thus, the Fr{\'e}chet bounds apply and gives us the bounds $\max\{0, \pi_i + \pi_j - 1\} \leq \pi_{ij} \leq \min\{ \pi_i, \pi_j\}$ \citep{Frechet1935}. Putting these together, we have for the LPMs 
\begin{equation}
    \max\{0, \pi_i + \pi_j - 1\} \leq \pi_{ij} \leq \pi_i \pi_j. \label{eq:pi_ij_bounds}
\end{equation}
Incorporating these bounds into the algorithm should improve the estimation of the second-order probabilities for LPM sampling. The second issue is related to the choice of $M$, the number of replications. Whereas higher $M$ is expected to provide better estimates, the computation time increases linearly with $M$.


\section{Proposed variance estimators based on estimation of second-order probabilities} \label{sec:method1}
For conceptualizing the sampling designs, it is helpful to define the following ratio
\begin{equation}
    r_{ij} = \dfrac{\pi_{ij}}{\pi_i \pi_j}. \label{eq:ratio_ij}
\end{equation}
For the local pivotal methods, $r_{ij} \in [0,1]$, with $r_{ij} = 0$ representing a pair $(i,j)$ that are never included in the same sample and $r_{ij} = 1$ denoting pairs $(i,j)$ that appear in the sample independently of each other. The dependence of $r_{ij}$ on the distance between points $(i,j)$ is shown in Table \ref{tab:LPM_rij}.

\begin{table}[ht]
\centering
\caption{Ratio $r_{ij}$ in LPM sampling} \label{tab:LPM_rij}
\begin{tabular}{c|c}
Pair $(i,j)$ & $r_{ij}$ \\ \hline
Very close & $\approx 0$ \\
Moderately close & $\in (0,1)$ \\
Far apart & $\approx 1$ \\ \hline
\end{tabular}
\end{table}

\noindent The unadjusted empirical estimate of the ratio over $M$ replications obtained by substituting $\widehat{\pi}^{\text{F}}_{ij}$ from (\ref{eq:emp_pi_ij_unb}) for $\pi_{ij}$ in \eqref{eq:ratio_ij} is:
\begin{equation}
    \widehat{r}_{ij}^{\text{F}} = \dfrac{C_{ij} + 1}{(M+1) \pi_i \pi_j}, \label{eq:r_est_unadj}
\end{equation}
yielding the following adjusted estimator
\begin{equation}
\widehat r_{ij}^{\text{adj}} = \min \left \{ \max \left( \frac{\pi_i + \pi_j - 1}{\pi_i \pi_j}, \ \widehat r_{ij}^{\text{F}} \right) \, , \, 1 \right \}, \label{eq:r_est_adj}
\end{equation}
applying \eqref{eq:pi_ij_bounds}.

Our proposed method for estimation of \eqref{eq:syg_var} and \eqref{eq:chaud_pop_var} involves the following steps outlined in Procedure \ref{proc1}:
\begin{procedure} \label{proc1}
Suppose we have a sample of size $n$ from a population of size $N$ obtained using LPM, with known values of auxiliary variables $\{\bm{X}_i\}_{i=1}^N$ and first-order inclusion probabilities $\{\pi_i\}_{i=1}^N$ for all population units, along with responses $\{Y_i : i \in \mathcal{S}\}$ observed for the sampled units.
\begin{itemize}
\item[(i)] Obtain $M$ independent Monte Carlo samples using the same sampling design (LPM), auxiliary variables $\bm{X}_i$, first-order inclusion probabilities $\pi_i$ for $i=1,\dots,N$, and distance measure used to generate the original sample $\mathcal{S}$. Use these samples to estimate $r_{ij}$ by $\widehat r_{ij}^{\text{adj}}$ from (\ref{eq:r_est_adj}) for all $\binom{n}{2}$ pairs $(i,j)$ in the actual sample $\mathcal{S}$. 
\item[(ii)] Substitute $r_{ij} = \widehat r_{ij}^{\text{adj}}$ in the lead term $\left(\pi_i\pi_j - \pi_{ij} \right)/\pi_{ij} = 1/r_{ij} - 1$ of \eqref{eq:syg_var} to estimate $\widehat{\var}_{\textrm{SYG}}(\overline{y})$. For the remainder of the manuscript, we will refer to this estimator as ``SYG-E''.
\item[(iii)] Substitute $\pi_{ij} = \widehat{\pi}_{ij}^{\textrm{adj}} = \widehat r_{ij}^{\text{adj}} \pi_i \pi_j$ in \eqref{eq:chaud_pop_var} to estimate $s^2_{\text{Chaud}}$. For the remainder of the manuscript, we will refer to this estimator as ``Chaud-E''.
\end{itemize}
\end{procedure}

In order to gauge how large $M$ has to be to obtain reliable estimates of (\ref{eq:syg_var}) and (\ref{eq:chaud_pop_var}) keeping the computational effort reasonable, we conduct a small simulation study reported in the following subsection.


\subsection{Preliminary simulation to determine the choice of the Monte Carlo iteration size $M$} \label{ss:motivating_sim_study}

We generate a synthetic finite population of size $N = 500$, the details of which are reported in Section \ref{sec:validating_sim_study}, where each unit has two auxiliary variables. An LPM1 sample of size $n=50$ is drawn from this population, setting the first-order inclusion probability of each unit equal to $n/N = 0.1$. After drawing the sample, we apply Procedure \ref{proc1} to compute the SYG-E and Chaud-E estimators using a fixed value of $M$. The Procedure is repeated $R$ times to obtain $R$ SYG-E estimates $\widehat{\var}_{\textrm{SYG}}^{(r)}(\overline{y})$ and $R$ Chaud-E estimates $s_{\text{Chaud}}^{2\,(r)}$ for $r=1, \ldots, R$. We compare the SYG-E estimates to the ``true'' variance of the HT means over a large ($10^7$) number of samples:
\begin{equation}
    \mathbb{V}_{\mathrm{HT}}  = \dfrac{1}{10^7-1}\sum_{i=1}^{10^7} \left(\overline{y}^{(i)} - \dfrac{1}{10^7}\sum_{r=1}^{10^7} \overline{y}^{(r)}\right)^2 \label{eq:true_ht_var}
\end{equation}
where $\overline{y}^{(i)}$ denotes the HT-mean from the $i^{th}$ sample. Specifically, we compute the relative bias (RB) and mean squared error (MSE):
\begin{equation}
    \mathrm{RB_{SYG-E}} = 100 \times \frac{\overline{\mathbb{V}}_{\textrm{SYG}} - \mathbb{V}_{\mathrm{HT}}}{\mathbb{V}_{\mathrm{HT}}}, \qquad \mathrm{MSE}_{\textrm{SYG-E}} = \dfrac{1}{R}\sum_{r=1}^R (\widehat{\var}_{\textrm{SYG}}^{(r)}(\overline{y}) - \mathbb{V}_{\mathrm{HT}})^2 \label{eq:RB_MSE}
\end{equation}
where $\overline{\mathbb{V}}_{\textrm{SYG}} = \sum_{r=1}^R \widehat{\var}_{\textrm{SYG}}^{(r)}(\overline{y})/R.$ The RB and MSE for the Chaud-E estimator are obtained in a similar manner by comparing the $R$ Chaud-E estimates with the true population variance:
\begin{equation}
    \mathbb{V}_{\mathrm{pop}} = \dfrac{1}{N-1}\sum_{i=1}^N \bigg( Y_i - \dfrac{1}{N}\sum_{j=1}^N Y_j \bigg)^2 \label{eq:true_pop_var}.
\end{equation}

\begin{table}[htbp]
    \centering
    \begin{tabular}{c|c|c|c|c}
        \toprule
        Estimator & $M$ & RB (\%) & MSE & Avg.\ Time (seconds) \\
        \midrule
        \multicolumn{5}{l}{Variance of HT Mean} \\
        \midrule
        \multirow{5}{*}{SYG-E}
            & $10^5$   & 416.67 & $1.55 \times 10^{-4}$ & 5.24     \\
            & $5\times10^5$   & 171.11 & $2.84 \times 10^{-5}$ & 25.94   \\
            & $10^6$   & 114.15 & $1.38 \times 10^{-5}$ & 51.78    \\
        \midrule
        \multicolumn{5}{l}{Population variance} \\
        \midrule
        \multirow{5}{*}{Chaud-E}
            & $10^5$   & 1.79   & $1.88 \times 10^{-2}$ &  5.24     \\
            & $5\times10^5$   & 1.10   & $1.84 \times 10^{-2}$ & 25.94    \\
            & $10^6$   & 0.94   & $1.83 \times 10^{-2}$ & 51.78   \\
    \bottomrule
    \end{tabular}
    \caption{Relative bias, MSE, and average runtime for variance of HT mean and population variance estimators. SYG-E and Chaud-E are shown at three Monte Carlo resolutions $M$. Reported times are estimates of how long the computation took to run the simulations with nine cores running in parallel.}
    \label{tab:syg_e_sim_study}
\end{table}

Table \ref{tab:syg_e_sim_study} reports the relative bias and mean squared error over the $R= 500$ iterations for SYG-E and Chaud-E across three resolutions $M \in \{10^5, 5 \times 10^5, 10^6\}$, along with the average time $M$ replications took. The performance of the Chaud-E estimator is found to be satisfactory even at the lowest $M$ ($10^5$), with a relative bias of 1.79\%. However, the SYG-E estimator, while having a relatively small variance, appears to have a very high positive relative bias, which is 114.15\% even at $M = 10^6$. To explore the reason behind such a huge bias, we try to understand the role of $M$ on the empirical estimation of the second-order probabilities.

\begin{figure}[ht]
    \centering
    \includegraphics[width=0.75\linewidth]{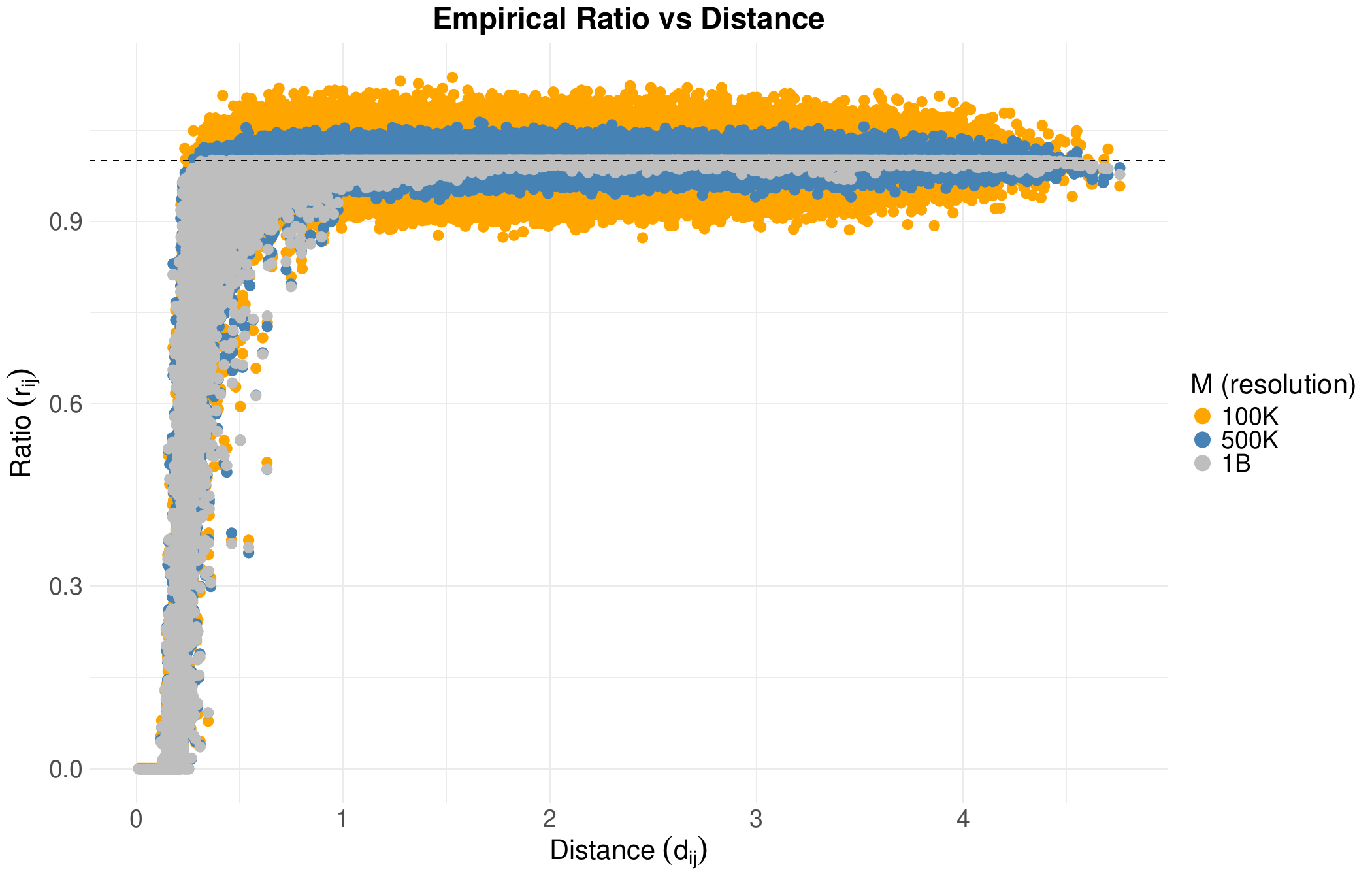}
    \caption{All $\binom{N}{2} = 124{,}750$ unadjusted empirical ratios $\widehat{r}_{ij}^{\textrm{F}}$ for this specific population and sample size at three different resolutions $M \in \{10^5, 5 \times 10^5, 10^9\}$ versus the distance between units $i$ and $j$. Synthetic population is $N = 500$, sample size $n= 50$, equal first-order inclusion probabilities, and using LPM1. Orange points are for $M = 10^5$, blue for $M = 5 \times 10^5$, and gray for $M = 10^9$.}
    \label{fig:emp_ratios}
\end{figure}

Figure \ref{fig:emp_ratios} shows all $\binom{N}{2} = 124{,}750$ unadjusted empirical ratios (\ref{eq:r_est_unadj}) at three different $M \in \{10^5, 5 \times 10^5, 10^9\}$ versus the distance between units $i$ and $j$. This figure helps reveal an interesting dependence of empirical estimation on $M$: at small distances, the increased number of replications $M$ does not dramatically change the estimates, but at large distances, as $M$ increases, the empirical ratios become more tightly concentrated around 1. 

For a single LPM sample, if we estimate all $\binom{n}{2}$ sampled pairs empirically using a reasonable $M$ such as $M = 5 \times 10^5$, the majority of the pairs will be in the moderate to large distance range. Within this range, the unadjusted empirical ratios $\widehat{r}_{ij}^{\textrm{F}}$ are noisy estimates of the true ratio, which is approximately 1 for independent pairs (as seen with the $M = 10^9$ estimates). The leading coefficient $(\pi_ i \pi_j - \pi_{ij})/\pi_{ij} = 1/r_{ij} - 1$ in \eqref{eq:syg_var} would be approximately zero for pairs in this distance range. However, with $M  \le 5 \times 10^5$, the adjusted empirical estimate $\widehat{r}^{\textrm{adj}}_{ij}$ (\ref{eq:r_est_adj}) fluctuates just below 1, leading the $(1/\widehat{r}^{\text{adj}}_{ij} - 1)$ term in the SYG estimate to be nonzero, creating artificial variance contributions from pairs that are effectively independent. Since most sampled pairs occur at moderate to large distances where $r_{ij} \approx 1$, the accumulation of these small errors across many pairs can dramatically inflate the variance estimate. Increasing $M$ to $10^9$ appears to be a solution to this overestimation problem, but is computationally infeasible, taking about 54379 seconds (about fifteen hours) in the current setting with only two auxiliary variables and a relatively small population size of 500.


\subsection{Modified procedure that incorporates a hypothesis test for independence between pairs} \label{ss:test_independence}

The results in Table \ref{tab:syg_e_sim_study} show that for this population SYG-E's accuracy improves as $M$ increases, but getting near zero relative bias comes at a substantial computational cost. Figure \ref{fig:emp_ratios} shows that at large distances the empirical ratios $r_{ij}$ concentrate around one, indicating that distant pairs behave as though being sampled independently. Motivated by these observations, we modify Procedure \ref{proc1} by (a) incorporating a test of the null hypothesis of independence of sampling indicators $S_i$ and $S_j$ and (b) setting the estimates of the empirical ratios $r_{ij}$ to one if the hypothesis cannot be rejected.

\begin{procedure} \label{proc2}
Suppose we have a sample of size $n$ from a population of size $N$ obtained using LPM, with known values of auxiliary variables $\{\bm{X}_i\}_{i=1}^N$ and first-order inclusion probabilities $\{\pi_i\}_{i=1}^N$ for all population units, along with responses $\{Y_i : i \in \mathcal{S}\}$ observed for the sampled units.
\begin{itemize}
\item[(i)] Obtain $M$ independent Monte Carlo samples using the same sampling design (LPM), auxiliary variables $\bm{X}_i$, first-order inclusion probabilities $\pi_i$ for $i=1,\dots,N$, and distance measure used to generate the original sample $\mathcal{S}$. Use these samples to estimate $r_{ij}$ by $\widehat r_{ij}^{\text{adj}}$ from (\ref{eq:r_est_adj}) for all $\binom{n}{2}$ pairs $(i,j)$ in the actual sample $\mathcal{S}$. 
\item[(ii)] For each pair $(i,j)$, test the null hypothesis $H_0^{ij}: r_{ij}=1$ against the alternative $H_1^{ij}: r_{ij} < 1$ at level of significance $\alpha$ (see Remark \ref{remark1}). If the test fails to reject the null hypothesis, set $\widehat r_{ij}^{\text{adj}}$ to 1. Else retain the estimated value.
\item[(iii)] Substitute $r_{ij} = \widehat r_{ij}^{\text{adj}}$ in the lead term $\left(\pi_i\pi_j - \pi_{ij} \right)/\pi_{ij} = 1/r_{ij} - 1$ of \eqref{eq:syg_var} to estimate $\widehat{\var}_{\textrm{SYG}}(\overline{y})$. For the remainder of the manuscript, we will refer to this estimator as ``SYG-P''.
\item[(iv)] Substitute $\pi_{ij} = \widehat{\pi}_{ij}^{\textrm{adj}} = \widehat r_{ij}^{\text{adj}} \pi_i \pi_j$ in \eqref{eq:chaud_pop_var} to estimate $s^2_{\text{Chaud}}$. For the remainder of the manuscript, we will refer to this estimator as ``Chaud-P''.
\end{itemize}
\end{procedure}

\begin{remark} \label{remark1}
To test the null hypothesis $H_0^{ij}: r_{ij}=1$, or equivalently, $\pi_{ij} = \pi_i \pi_j$, we use the test statistic $C_{ij}$, the number of times units $i$ and $j$ were sampled together in $M$ independent replications of LPMs on the full population. Under $H_0^{ij}$, $C_{ij}$ is a binomial random variable with parameters $M$ and $\pi_i \pi_j$. Thus the exact $p$-value for the alternative $H_1^{ij}: r_{ij}<1$, or equivalently, $\pi_{ij} < \pi_i \pi_j$ is:
\begin{equation}
    p_{ij} = \Pr\big(C_{ij} \leq c_{ij} \mid \pi_{ij} = \pi_i \pi_j \big) = \sum_{k=0}^{c_{ij}} \binom{M}{k} (\pi_i \pi_j)^k (1-\pi_i\pi_j)^{M-k}, \label{eq:c_ij_p_value}
\end{equation}
where $c_{ij}$ denotes the observed value of $C_{ij}$ for sampled pair $(i,j)$. The null hypothesis is rejected at level $\alpha$ if $p_{ij} \le \alpha$.
\end{remark}

\begin{remark} \label{remark2}
Since Procedure \ref{proc2} includes a hypothesis test for each of the $\binom{n}{2}$ sampled pairs, adopting a multiple testing procedure like the Bonferroni correction to control the probability of Type-I error seems natural. However, in the current problem, each type-I error results in plugging in a positive value of the term $(\pi_i \pi_j - \pi_{ij})/\pi_{ij}$ in (\ref{eq:syg_var}) instead of its true value of zero, consequently resulting in overestimation of the variance. On the contrary, a \emph{statistically conservative} procedure like the Bonferroni correction reduces the power of the test, which means several positive terms $(\pi_i \pi_j - \pi_{ij})/\pi_{ij}$ in (\ref{eq:syg_var}) being replaced by zeros, leading to underestimation of the variance. Thus, in this setting, to avoid underestimating variances, we prefer to incorrectly retain empirical estimates of $\pi_{ij}$ for many pairs, rather than incorrectly classifying dependent pairs as independent, and therefore do not adopt a multiple-testing correction.
\end{remark}


\section{Proposed variance estimators based on model-assisted bootstrap (MABS)} \label{sec:method2}

Previous literature has utilized a set of observations with both response and auxiliary information to leverage additional units where only auxiliary information is available \citep{rubin1987, kim2021, mozer2024}. Similarly, our second approach departs from the need for second-order inclusion probabilities entirely. Rather than estimating $\pi_{ij}$ directly, we fit a predictive model to the sampled units and use it to construct a synthetic population from which we resample to estimate the variance of the HT mean and population variance.

We first construct a predictor for response $Y$ from the observed data $\{(\bm{X}_i, Y_i) : i \in \mathcal{S}\}$. Since the vector of auxiliary variables $\mathbf{X}_i$ is available for all units $i=1, \ldots, N$ of the population, the fitted model can be used to obtain a prediction $\widehat{Y}_i$ of $Y_i$ for each population unit $i=1, \ldots, N$. A natural option is to use an \emph{interpolating} predictor satisfying $\widehat{Y}_i = Y_i$ for $i \in \mathcal{S}$, i.e., the predicted response values for sampled units exactly match their observed values. We describe the proposed method for creating such a predictor in Section \ref{ss:GP-model}. 

Once a synthetic population $\{\widehat{Y}_i\}_{i=1}^N$ is constructed, we repeatedly draw bootstrap samples $\mathcal{S}^{(1)}, \ldots, \mathcal{S}^{(B)}$ from this population using the same sampling design (LPM), auxiliary variables $\mathbf{X}_i$ and first-order inclusion probabilities $\pi_i$ for $i=1, \ldots, N$ used to generate the original sample $\mathcal{S}$. For each bootstrap sample, the HT mean is computed and the empirical variance across the $B$ bootstrap samples provides an estimate of the variance of the HT mean. The proposed MABS method is outlined in Procedure \ref{proc3}.

\begin{procedure} \label{proc3}
Suppose we have a sample of size $n$ from a population of size $N$ obtained using LPM, with known values of auxiliary variables $\{\bm{X}_i\}_{i=1}^N$ and first-order inclusion probabilities $\{\pi_i\}_{i=1}^N$ for all population units, along with responses $\{Y_i : i \in \mathcal{S}\}$ observed for the sampled units.

\begin{itemize}
\item[(i)] Fit an interpolating prediction model for response $Y$ from the observed data $\{(\bm{X}_i, Y_i) : i \in \mathcal{S}\}$ that satisfies $\widehat{Y}_i = Y_i$ for $i \in \mathcal{S}$ following the methodology outlined in Sec \ref{ss:GP-model}.
\item[(ii)] Apply the fitted predictor to $\bm{X}_i$ for all $i=1,\dots, N$ to construct the synthetic population $\{\widehat Y_i\}_{i=1}^N$, where the interpolation property ensures that $\widehat{Y}_i = Y_i$ for $i \in \mathcal{S}$.
\item[(iii)] Obtain bootstrap samples $\mathcal{S}^{(1)}, \ldots, \mathcal{S}^{(B)}$ from the synthetic population using the same sampling design (LPM), auxiliary variables $\mathbf{X}_i$, first-order inclusion probabilities $\pi_i$ for $i=1, \ldots, N$, and distance measure used to generate the original sample $\mathcal{S}$.
\item[(iv)] For bootstrap sample $b = 1, \ldots, B$, compute the HT-mean
\begin{equation}
    \overline{y}_{\mathrm{pred}}^{(b)} = \dfrac{1}{N}\sum_{i \in \mathcal{S}^{(b)}} \dfrac{\widehat{Y}_i}{\pi_i}. \label{eq:surrogate_ht_mean}
\end{equation}
\item[(v)] Estimate the variance of the HT mean by the empirical variance across the $B$ bootstrap replicates, calling it the Model-Assisted BootStrap (MABS) estimator:
\begin{equation}
    \widehat{\var}_{\mathrm{MABS}}(\overline{y}) = \dfrac{1}{B-1}\sum_{b=1}^B \bigg( \overline{y}_{\mathrm{pred}}^{(b)} - \dfrac{1}{B}\sum_{r=1}^B \overline{y}_{\mathrm{pred}}^{(r)} \bigg)^2. \label{eq:ht_var_MABS}
\end{equation}
\item[(vi)] Finally, estimate the population variance
\begin{equation}
    s_{\mathrm{MABS}}^2 = \frac{1}{N-1}\sum_{i=1}^N \left(\widehat{Y}_i - \overline{\widehat{Y} }\right)^2, \label{eq:pop_var_MABS}
\end{equation}
where $\overline{\widehat{Y} } = \sum_{i=1}^N \widehat{Y}_i / N$ is the mean of the $N$ predicted values.
\end{itemize}
\end{procedure}


\subsection{Building an interpolating predictor from the sampled units} \label{ss:GP-model}

To obtain an interpolating predictor from the observed data $\{(\bm{x}_i, Y_i) : i \in \mathcal{S}\}$, we use a Gaussian process (GP) model, in which the response function $Y(\mathbf{x})$ at $\mathbf{X} = \mathbf{x}$ is assumed to be of the form
\begin{equation}
    Y(\mathbf{x}) = \mu_0 + Z(\mathbf{x}) + W(\mathbf{x}), \label{eq:gp_model}
\end{equation}
where $\mu_0$ is the mean of the response function and $Z(\mathbf{x})$ is a Gaussian process with mean 0 and covariance function 
$$\cov\left( Z(\mathbf{x}), Z(\mathbf{x}^{\prime})\right) = \sigma^2 R(\mathbf{x}, \mathbf{x}^{\prime})$$ 
for input points $(\mathbf{x}, \mathbf{x}^{\prime})$. The final term in \eqref{eq:gp_model}, $W(\mathbf{x})$, is a zero-mean Gaussian process with covariance function
$$ \cov\left( W(\mathbf{x}), W(\mathbf{x}^{\prime})\right) = \tau^2 \  \mathbb{I}(\mathbf{x} = \mathbf{x}^{\prime}),$$ for input points $(\mathbf{x}, \mathbf{x}^{\prime})$ 
where $\mathbb{I}(\cdot)$ denotes the indicator function. $Z(\bm{x})$ and $W(\bm{x})$ are assumed to be independent.  

Several choices of $R(\bm{x}, \bm{x}^{\prime})$ are possible. Here we propose using a Mat\'ern covariance function with smoothness parameter $\nu = 5/2$ \citep{Santner2003, RasmussenWilliams2006}. Under this choice, the structured GP component $Z(\bm{x})$ is twice mean-square differentiable, providing a sufficiently smooth component for prediction while the nugget term accommodates unstructured variation not captured by the auxiliary variables.

To estimate the unknown parameters of the GP model and subsequently obtain the predictor, we use the ordinary Kriging method \citep{Santner2003} with an estimated nugget variance $\tau^2$. All model parameters are estimated by maximum likelihood on the sampled units via the \texttt{km()} function in the \texttt{DiceKriging} \texttt{R} package \citep{DiceKriging}. Conditional on the fitted GP model and observed sample, the Kriging predictive distribution for each population unit $i$ with auxiliary variables $\bm{X}_i$ is Gaussian with posterior predictive mean $\mu_i$ and posterior predictive variance $\sigma_i^2$ \citep{RasmussenWilliams2006}. Under the nugget-effect formulation, prediction interpolates the observed responses, so that $\mu_i = Y_i$ and $\sigma_i = 0$ for all $i \in \mathcal{S}$. Instead of using the posterior mean $\mu_i$ directly as our prediction, we generate a synthetic response for each population unit by drawing a single realization from its marginal posterior predictive distribution:
\begin{equation}
    \widehat{Y}_i \sim \mathrm{N} (\mu_i \, ,\, \sigma_i^2), \qquad i = 1, \dots, N. \label{eq:mabs_marginal}
\end{equation}

This model-fitting approach has two advantages, making it appropriate for our problem. First, the sampled outcomes are trusted and retained exactly, because the conditional predictions satisfy $\widehat{Y}_i = Y_i$,
and have zero conditional variance at sampled units. Second, the non-sampled units retain uncertainty from both the structured GP component and the unit-specific nugget component. The latter cannot be inferred from outcomes at distinct units and therefore remains in the predictive variance.


\section{Simulation Study} \label{sec:validating_sim_study}

Throughout the simulation studies in this section, we consider a single synthetic finite population of $N = 500$ units. To generate an auxiliary space, we drew a Latin hypercube sample \citep{mckay1979} of $N = 500$ from $[0,1]^2$ using the \texttt{randomLHS()} function in the \texttt{R} package \texttt{lhs} \citep{lhsR}, yielding auxiliary vectors $\bm{x}_i = (x_{1i}, x_{2i}) \in [0,1]^2$ for $i=1,\dots, N$. The response $Y_i \in \mathbb{R}$ for each unit is the result of plugging $\bm{x}_i$ into the rescaled Branin function and adding a small Gaussian noise term:
\begin{equation}
    Y_i = \dfrac{1}{51.27}\bigg[ \bigg(\overline{x}_{2i} - \dfrac{5.1}{(4\pi^2)}\overline{x}_{1i}^2 + \dfrac{5}{\pi} \overline{x}_{1i} - 6 \bigg)^2 + \bigg(10 - \dfrac{10}{(8 \pi)} \bigg) \cos(\overline{x}_{1i}) - 44.31 \bigg] + \epsilon_i \label{eq:branin}
\end{equation}
with $\overline{x}_{1i} = 15x_{1i} - 5$, $\overline{x}_{2i} = 15x_{2i}$ \citep{simulationlib}, and $\epsilon_i \overset{iid}{\sim}\mathrm{N}(0, \eta^2)$. Unless specified otherwise, we take $\eta = 0.2$, marking a small amount of noise being added to the response variables. The rescaled Branin function is a standard test function in computer experiments with mean zero and standard deviation one, chosen here because it exhibits nonlinear spatial structure across $[0,1]^2$, making the spatial balance in auxiliary space consequential in estimation.


\subsection{LPM1 on single synthetic population} \label{sec:single_pop_LPM1}
We evaluate our proposed variance estimators over $R = 2{,}500$ iterations on the synthetic Branin population described earlier, with $N = 500$, $n=50$, equal first-order inclusion probabilities, Euclidean distance, and LPM1. At each iteration, $M = 5 \times 10^5$ Monte Carlo replications are used to estimate the second-order inclusion probabilities of the $\binom{n}{2} = 1225$ sampled pairs. For the SYG-P estimates obtained using Procedure \ref{proc2}, we use the $M = 5\times 10^5$ empirical estimates and significance level $\alpha = 0.05$, while the MABS estimator obtained using Procedure \ref{proc3} uses the GP model and $B = 2{,}000$ bootstrap samples. 

\begin{figure}[ht]
    \centering
    \includegraphics[width=0.75\linewidth]{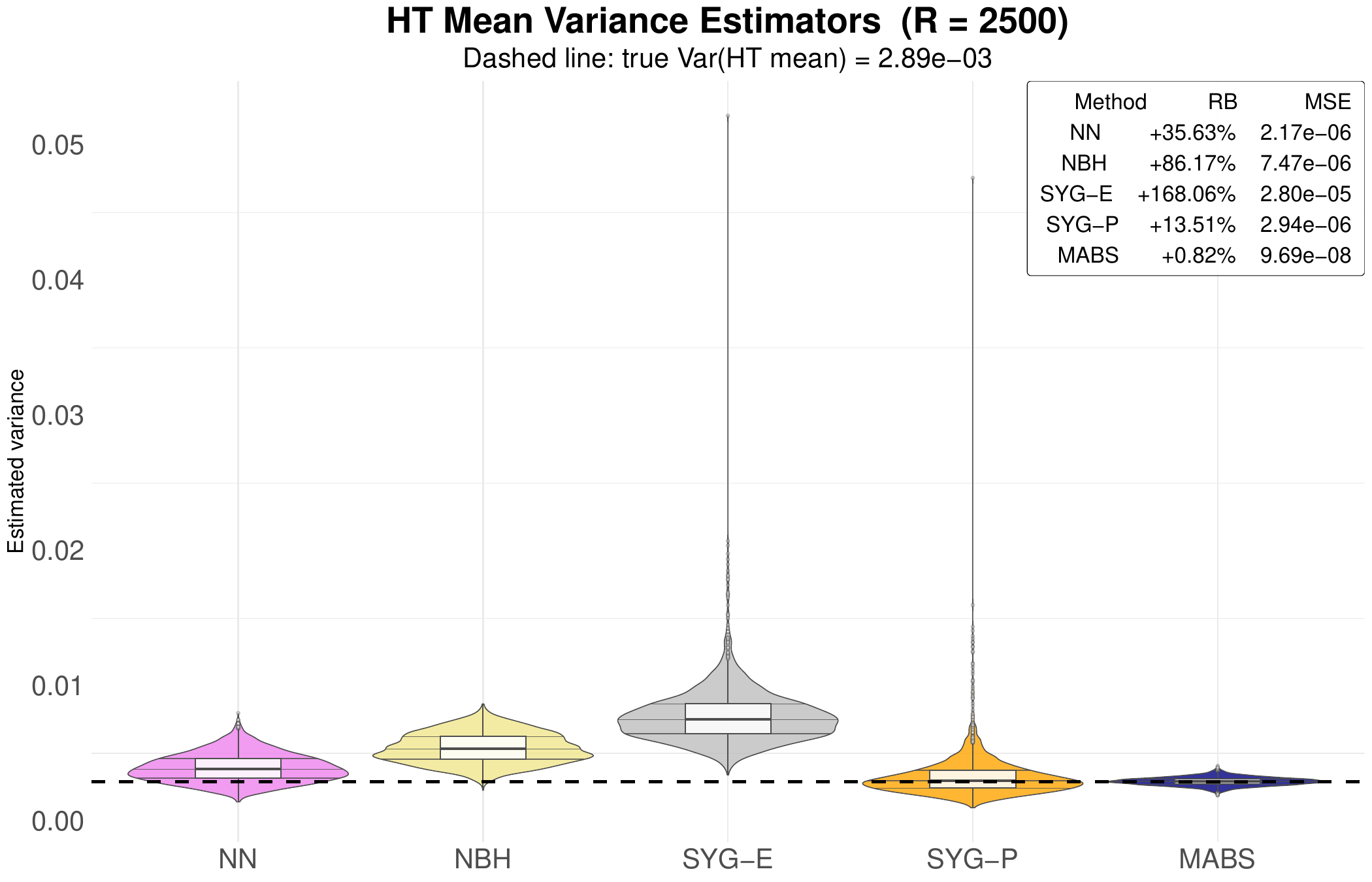}
    \caption{Violin plots of $R = 2{,}500$ variance estimates using five different methods to estimate the variance of HT mean: NN (\ref{eq:near_neigh}), NBH (\ref{eq:NBH_var}), SYG-E (\ref{eq:syg_var}) with the $M = 5 \times 10^5$ empirical estimates, SYG-P obtained using Procedure \ref{proc2} with $\alpha = 0.05$, and MABS (\ref{eq:ht_var_MABS}) obtained using Procedure \ref{proc3} with $B = 2{,}000$ bootstrap LPM1 samples. The horizontal line is at the ``true'' variance (\ref{eq:true_ht_var}).}
    \label{fig:ht_var_sd02_n50}
\end{figure}

\begin{figure}[ht]
    \centering
    \includegraphics[width=0.75\linewidth]{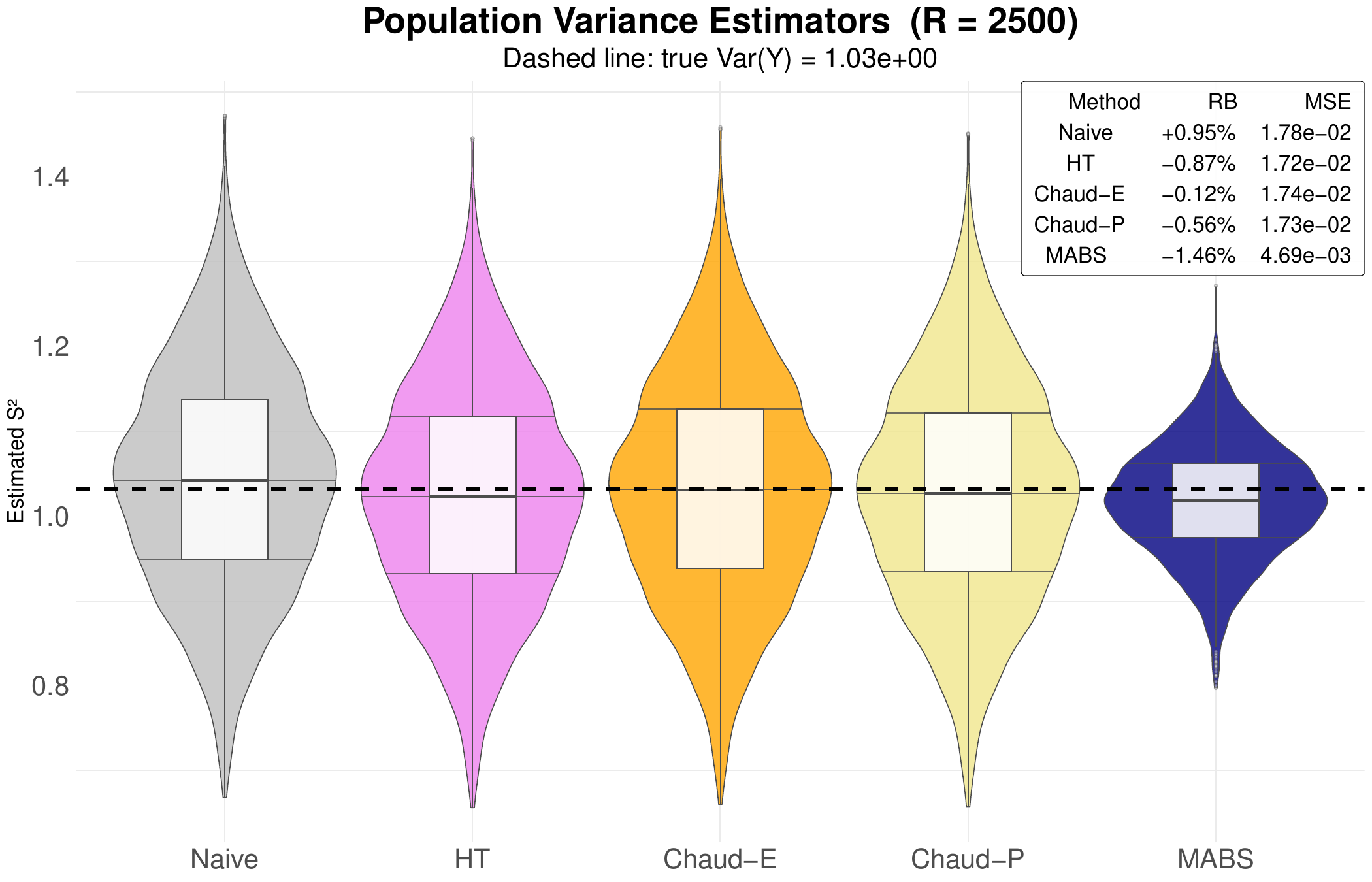}
    \caption{Violin plots of $R = 2{,}500$ variance estimates using five different methods to estimate the population variance: Naive (\ref{eq:naive_pop_var}), HT (\ref{eq:ht_pop_var}), Chaud-E obtained using Procedure \ref{proc1} with the $M = 5 \times 10^5$ empirical estimates, Chaud-P obtained using Procedure \ref{proc2} with $\alpha = 0.05$, and MABS estimator obtained using Procedure \ref{proc3}. The horizontal line is at the ``true'' variance (\ref{eq:true_pop_var}).}
    \label{fig:pop_var_sd02_n50}
\end{figure}

Figure \ref{fig:ht_var_sd02_n50} shows the distributions of each different estimator of the variance of the HT mean. The SYG-P estimator resulted in the second closest to zero relative bias and mean squared error comparable to the NN estimator. The MABS estimator resulted in a relative bias of only $0.82\%$ while having a significantly lower mean squared error than the other estimators. Since this synthetic population is defined by the rescaled Branin function (\ref{eq:branin}) with small Gaussian noise $\eta = 0.2$, the correlations between the auxiliary and response variables are very strong, so the GP model is able to construct a highly accurate surrogate population. We will show in Section \ref{sec:sim_study_across_noise_levels} that as $\eta$ grows, the MABS estimator continues to have the best, or second best, relative bias and mean squared error. However, as $\eta$ grows, the SYG-P begins to substantially underestimate the variance. 

Figure \ref{fig:pop_var_sd02_n50} shows the distributions of the each different estimate of population variance. The MABS estimator of the variance has the smallest mean squared error, whereas the other four methods have very similar mean squared errors. The MABS estimator has the largest relative bias of the five, followed by the naive estimator (\ref{eq:naive_pop_var}) estimator, although both relative biases are quite small (-1.46\% and +0.95\% respectively). The Chaud-E estimator is the closest relative bias to zero, in Section \ref{sec:sim_study_across_noise_levels}, we will see its performance stays stable as $\eta$ increases.


\subsection{Comparison at different noise levels}\label{sec:sim_study_across_noise_levels}
Before making any suggestions of which estimators to use, it is important to explore the performance of the estimators beyond the single finite population tested in the previous section. To examine how estimator performance changes as the relationship between the auxiliary and response variables becomes weaker, we consider four noise levels $\eta \in \{0.2, 0.4, 0.8, 1.2\}$. Specifically, we generate $\xi_i \overset{iid}{\sim}\mathrm{N}(0,1)$ once for each $i=1,\dots, N$ and define the noise contribution to unit $i$ as $\eta \times\xi_i$. The same realization $\{\xi_i\}_{i=1}^N$ is used at each noise level, so that increasing $\eta$ changes only the magnitude of the noise while preserving its underlying realization across population units. This allows differences in estimator performance across noise levels to be attributed more directly to the weakening relationship between the auxiliary and response variables, rather than to differences arising from separate random realizations of the noise. For the three additional levels $\eta \in \{0.4, 0.8, 1.2\}$, we run $R = 2{,}500$ iterations of variance estimation and compare the same estimators as in Figures \ref{fig:ht_var_sd02_n50} and \ref{fig:pop_var_sd02_n50}.

\begin{table}[ht]
    \centering
    \small   \begin{tabular}{l rr rr rr rr}
        \toprule
        & \multicolumn{2}{c}{$\eta = 0.2$} & \multicolumn{2}{c}{$\eta = 0.4$}
        & \multicolumn{2}{c}{$\eta = 0.8$} & \multicolumn{2}{c}{$\eta = 1.2$} \\
        \cmidrule(lr){2-3} \cmidrule(lr){4-5} \cmidrule(lr){6-7} \cmidrule(lr){8-9}
        Estimator & RB & RRMSE & RB & RRMSE & RB & RRMSE & RB & RRMSE \\
        \midrule
        NN      &   35.63 &  50.92 &  26.30 &  40.19 &  18.60 & 34.90 & 14.81 & 32.72 \\
        NBH     &   86.17 &  94.50 &  43.75 &  51.17 &   5.37 & 20.12 & $-$6.15 & 18.95 \\
        SYG-E   &  168.06 & 183.08 &  89.82 & 114.66 &  23.29 & 60.77 &  5.08 & 64.31 \\
        SYG-P   &   13.51 &  59.32 & $-$8.15 & 65.82 & $-$26.74 & 59.67 & $-$31.04 & 69.75 \\
        MABS    & 0.82 &  10.76 & 5.42 & 18.19 & 7.69 & 24.52 & 6.39 & 26.50 \\
        \bottomrule
    \end{tabular}
    \caption{Relative bias (RB, \%) and relative root mean squared error (RRMSE, \%) of the NN, NBH, SYG-E, SYG-P, and MABS estimators of $\mathbb{V}_{\mathrm{HT}}$, at four noise levels $\eta$. Larger $\eta$ means a weaker relationship between the auxiliary and response variables. Results are over $R = 2500$ iterations with $N = 500$, $n = 50$, and the LPM1 design using equal inclusion probabilities.}
    \label{tab:ht_var_n50_across_noises}
\end{table}

Table \ref{tab:ht_var_n50_across_noises} reports percentage relative bias (RB) defined in \eqref{eq:RB_MSE} and the relative root mean squared error (RRMSE) defined as
\begin{equation}
    \mathrm{RRMSE_{SYG-E}} = 100 \times \dfrac{\sqrt{\mathrm{MSE_{SYG-E}}}}{\mathbb{V}_{\mathrm{HT}}}, \label{eq:relative_root_mse}
\end{equation} 
where $\mathbb{V}_{\mathrm{HT}}$ and $\mathrm{MSE_{SYG-E}}$ are from (\ref{eq:true_ht_var}) and (\ref{eq:RB_MSE}), respectively. These metrics are shown for the five different estimators of the variance of the HT mean compared in Figure \ref{fig:ht_var_sd02_n50} across four different noise levels $\eta$. Clearly the MABS estimator exhibits the most robust performance with respect to both criteria across the different noise levels and dominates the other four estimators almost at all noise levels. The performance of the SYG-P estimator, while promising for the smallest noise level $\eta = 0.2$, deteriorates sharply as $\eta$ increases and ends up with a large negative relative bias when $\eta = 1.2$.

Table \ref{tab:pop_var_n50_across_noises} reports the RB and the RRMSE of the five different estimators of the population variance compared in Figure \ref{fig:pop_var_sd02_n50} across the four noise levels. As in the case of the lowest noise, none of these methods appear to have a significant advantage over the others in terms of RB and RRMSE.

As noted in Remark \ref{remark_pi_ij_zero}, pairs with $\pi_{ij} = 0$ never contribute their corresponding contrasts to the SYG and Chaudhuri estimators. At low noise levels, these pairs tend to be very close in the auxiliary space and therefore also have similar responses, making the omitted differences relatively small. As $\eta$ increases, however, nearby units can have increasingly different responses, so the unobservable contrasts associated with pairs having $\pi_{ij} = 0$ become larger and the resulting downward bias grows. At the same time, the small deviations of $\widehat r_{ij}^{\text{adj}}$ from one for effectively independent pairs at moderate to large distances, as explored in Figure \ref{fig:emp_ratios}, introduce upward bias through artificial contributions to the SYG estimator. At low noise levels, this source of overestimation can be substantial relative to the downward bias from the omitted differences. SYG-P performs well in this setting because its snapping correction removes many of the Monte Carlo-induced contributions. As the noise level increases, the downward bias from the unobservable contrasts becomes more consequential, while the SYG-P does not correct for this source of bias. This provides a plausible explanation for its deterioration as $\eta$ increases. Although the Chaud-E and Chaud-P estimators are subject to the same issue when $\pi_{ij} = 0$, the results in Table \ref{tab:pop_var_n50_across_noises} suggest their performance is substantially less affected by it in the settings considered here. 

\begin{table}[htbp]
    \centering
    \small
    \begin{tabular}{l rr rr rr rr}
        \toprule
        & \multicolumn{2}{c}{$\eta = 0.2$} & \multicolumn{2}{c}{$\eta = 0.4$}
        & \multicolumn{2}{c}{$\eta = 0.8$} & \multicolumn{2}{c}{$\eta = 1.2$} \\
        \cmidrule(lr){2-3} \cmidrule(lr){4-5} \cmidrule(lr){6-7} \cmidrule(lr){8-9}
        Estimator & RB & RRMSE & RB & RRMSE & RB & RRMSE & RB & RRMSE \\
        \midrule
        Naive   &  0.95 & 12.92 &  1.25 & 14.38 &  0.61 & 16.64 &  0.90 & 17.88 \\
        HT      &  $-$0.87 & 12.69 & $-$0.58 & 14.08 & $-$1.20 & 16.38 & $-$0.92 & 17.56 \\
        Chaud-E &  $-$0.12 & 12.77 &  0.26 & 14.19 &  $-$0.14 & 16.49 &  0.33 & 17.75 \\
        Chaud-P &  $-$0.56 & 12.72 & $-$0.17 & 14.12 & $-$0.57 & 16.42 & $-$0.10 & 17.66 \\
        MABS    & $-$1.46 & 6.63 & $-$2.04 & 11.41 & $-$2.03 & 16.40 & $-$1.03 & 18.31 \\
        \bottomrule
    \end{tabular}
    \caption{Relative bias (RB, \%) and relative root mean squared error (RRMSE, \%) of the Naive, HT, Chaud-E, Chaud-P, and MABS estimators of $\mathbb{V}_{\mathrm{pop}}$, at four noise levels $\eta$. Results are over $R = 2500$ iterations with $N = 500$, $n = 50$, and the LPM1 design using equal inclusion probabilities.}
    \label{tab:pop_var_n50_across_noises}
\end{table}


\subsection{LPM2 and using a non-Gaussian process model} \label{sec:sim_study_LPM2}

Having observed from the previous two subsections that the MABS estimator of the sampling variance of the HT mean outperforms its competitors over a range of signal-to-noise ratios, we now test its robustness to the noise distribution itself. Rather than Gaussian noise, we use $\epsilon_i \overset{iid}{\sim} \mathrm{Uniform}(-\sqrt{3}\eta, \, \sqrt{3} \eta)$ with $\eta = 0.2$. This uniform noise has mean zero and standard deviation $\eta = 0.2$.

Further, for this final test, instead of LPM1, we use LPM2, the faster and greedier version of LPM. As in the previous subsections, we evaluate the variance estimators over $R = 2{,}500$ iterations on the synthetic Branin population.    

Under this setting, we compare three estimators of the sampling variance of the HT-estimator: (i) the MABS estimator proposed and outlined in Procedure \ref{proc3}, (ii) a version of the MABS estimator where a random forest model is used to build a prediction model for out of sample responses instead of the GP-model described in Section \ref{ss:GP-model}, and (iii) the NN-estimator. 

\begin{figure}
    \centering
    \includegraphics[width=0.75\linewidth]{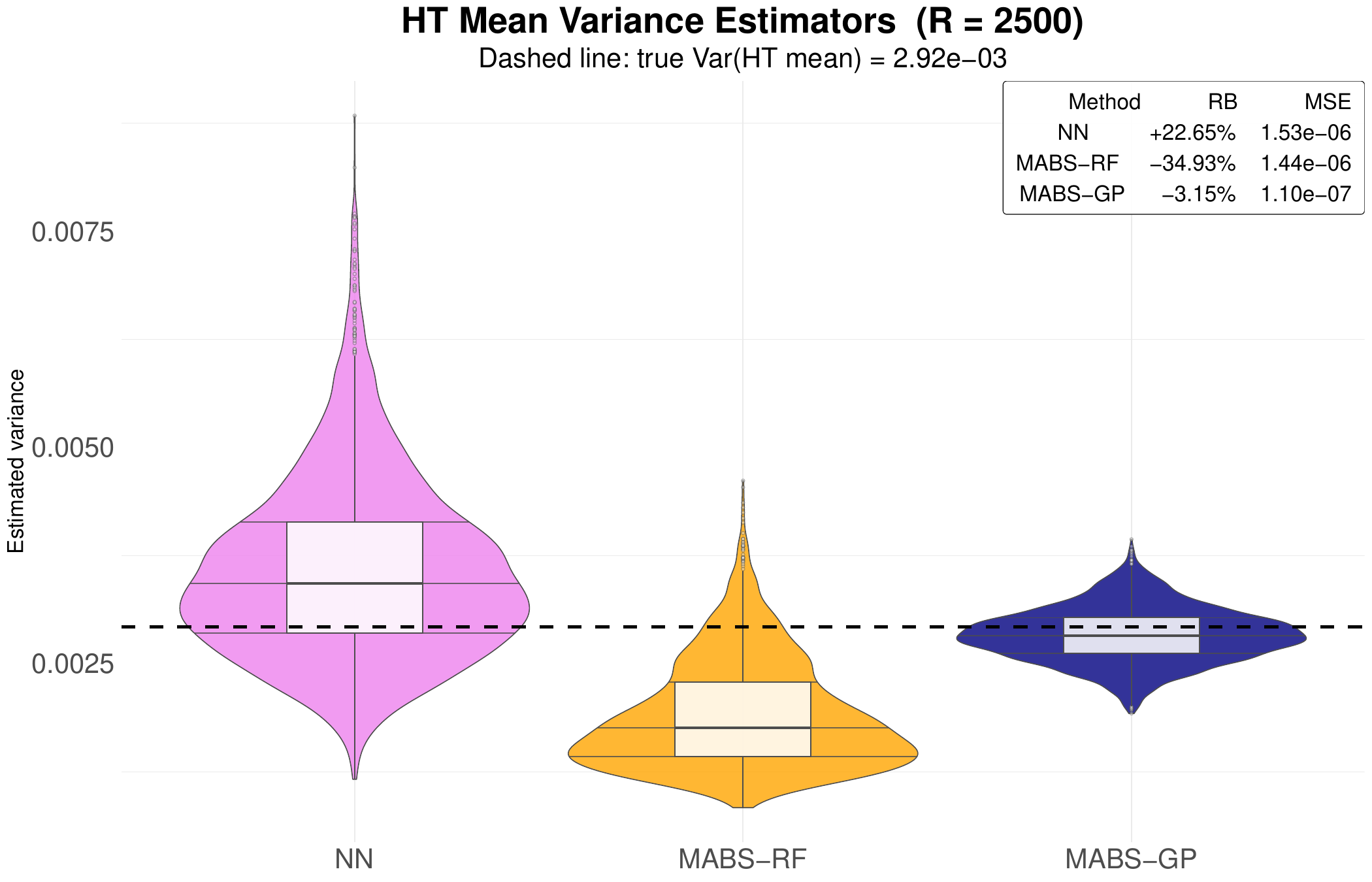}
    \caption{Violin plots of $R = 2{,}500$ variance estimates using three different methods to estimate the variance of HT mean: NN (\ref{eq:near_neigh}), MABS-GP (\ref{eq:ht_var_MABS}), and MABS-RF which is (\ref{eq:ht_var_MABS}) but the model fit is a random forest. The sampling design used is LPM2 with equal first-order inclusion probabilities. Both MABS estimators use $B = 2{,}000$ bootstrap LPM2 samples. The horizontal line is at the ``true'' variance (\ref{eq:true_ht_var}).}
    \label{fig:lpm2}
\end{figure}

Figure \ref{fig:lpm2} shows the distributions of the NN estimator as well as the two MABS estimators. Clearly, the proposed MABS estimator that uses GP modeling results in near zero relative bias and outperforms the other two estimators. The MABS estimator that uses random forest has a large negative bias. These results suggest that the proposed MABS estimator works well for LPM2 and non-Gaussian noise structures and reinforces the importance of using a GP model for prediction. Random forests, like any model that targets only the conditional mean $\mathbb{E}[Y_i \mid \bm{X}_i]$, assigns each population unit a smoothed value determined entirely by its auxiliary variables. The resulting surrogate population omits variation in $Y$ that the auxiliary variables do not explain, which is genuine variation in the finite population whenever the response and auxiliary variables do not have a one-to-one relationship. A realistic surrogate must reproduce the conditional distribution of $Y$ given $\bm{X}$, not only its mean. This motivates drawing each synthetic response from its posterior predictive distribution as outlined in Section \ref{ss:GP-model}.


\section{Discussion} \label{sec:discussion}

Spatially balanced sampling designs, including the local pivotal methods (LPM), can improve the precision of finite-population inference, but variance estimation remains challenging when second-order inclusion probabilities are unavailable or difficult to estimate. This paper proposes two approaches to this problem: a modified empirical inclusion-probability method and a model-assisted bootstrap (MABS) procedure based on an interpolating Gaussian process. The simulation results suggest that the methods can provide useful alternatives to existing approaches, with complementary performance across the settings considered. In particular, SYG-P performs well when the response is strongly associated with the auxiliary variables, while MABS exhibits comparatively stable performance for estimating the variance of the Horvitz–Thompson mean as the noise level increases. The results for estimating the finite-population variance are more mixed, and do not indicate a uniform advantage for MABS. The Chaud-P was a reliable estimate of the population variance within our tested synthetic populations, but further populations need to be tested to ensure the downward bias introduced by pairs with $\pi_{ij} = 0$ remains small across many populations. 

There are several directions for future research beyond the scope of this paper. An important one is to investigate the performance of the proposed methods in more varied finite populations and sampling settings, including higher-dimensional auxiliary spaces and more complex relationships between the response and auxiliary variables. Further work could also examine alternative predictive models within the MABS framework and develop a broader theoretical understanding of the properties of the proposed estimators. Finally, applications to real finite populations would help assess their practical performance and computational demands.


\bibliography{ref}

@article{grafstrom2012,
    author = {Grafstr{\"{o}}m, Anton and Lundstr{\"{o}}m, Niklas L. P. and Schelin, Lina},
    title = {Spatially Balanced Sampling Through the Pivotal Method},
    journal = {Biometrics},
    volume = {68},
    number = {2},
    pages = {514--520},
    year = {2012},
    month = {06},
    issn = {0006-341X},
    doi = {10.1111/j.1541-0420.2011.01699.x},
    url = {https://doi.org/10.1111/j.1541-0420.2011.01699.x},
    eprint = {https://academic.oup.com/biometrics/article-pdf/68/2/514/53237493/biometrics_68_2_514.pdf},
}

@article{neyman1934,
    author = {Neyman, Jerzy},
    title = {On the Two Different Aspects of the Representative Method : The Method of Stratified Sampling and the Method of Purposive Selection},
    journal = {Journal of the Royal Statistical Society},
    volume = {97},
    number = {4},
    pages = {558--606},
    year = {1934},
    month = {07},
    issn = {0952-8385},
    doi = {10.1111/j.2397-2335.1934.tb04184.x},
    url = {https://doi.org/10.1111/j.2397-2335.1934.tb04184.x},
    eprint = {https://academic.oup.com/jrsssa/article-pdf/97/4/558/49706846/jrsssa_97_4_558.pdf},
}

@misc{lin2022,
      title={Latin Hypercubes and Space-filling Designs}, 
      author={C. Devon Lin and Boxin Tang},
      year={2022},
      eprint={2203.06334},
      archivePrefix={arXiv},
      primaryClass={stat.ME},
      url={https://arxiv.org/abs/2203.06334}, 
}

@article{minasny2006,
    author = {Budiman Minasny and Alex B. McBratney},
    title = {A conditioned Latin hypercube method for sampling in the presence of ancillary information},
    journal = {Computers \& Geosciences},
    volume = {32},
    number = {9},
    pages = {1378--1388},
    year = {2006},
    issn = {0098-3004},
    doi = {10.1016/j.cageo.2005.12.009},
    url = {https://www.sciencedirect.com/science/article/pii/S009830040500292X}
}

@article{deville2004,
    author = {Jean-Claude Deville and Yves Till{\'{e}}},
    title = {Efficient Balanced Sampling: The Cube Method},
    year = {2004},
    ISSN = {00063444},
    URL = {http://www.jstor.org/stable/20441151},
    journal = {Biometrika},
    number = {4},
    pages = {893--912},
    publisher = {[Oxford University Press, Biometrika Trust]},
    urldate = {2026-10-03},
    volume = {91}
}

@article{grafstrom2013localcube,
    author = {Grafstr{\"{o}}m, Anton and Till{\'{e}}, Yves},
    title = {Doubly balanced spatial sampling with spreading and restitution of auxiliary totals},
    journal = {Environmetrics},
    volume = {24},
    number = {2},
    pages = {120--131},
    doi = {10.1002/env.2194},
    url = {https://onlinelibrary.wiley.com/doi/abs/10.1002/env.2194},
    eprint = {https://onlinelibrary.wiley.com/doi/pdf/10.1002/env.2194},
    year = {2013}
}

@article{horvitz1952,
 ISSN = {01621459, 1537274X},
 URL = {http://www.jstor.org/stable/2280784},
 author = {D. G. Horvitz and D. J. Thompson},
 journal = {Journal of the American Statistical Association},
 number = {260},
 pages = {663--685},
 publisher = {[American Statistical Association, Taylor & Francis, Ltd.]},
 title = {A Generalization of Sampling Without Replacement From a Finite Universe},
 urldate = {2025-05-27},
 volume = {47},
 year = {1952}
}

@article{sen1953,
  author  = {Sen, A. R.},
  title   = {On the Estimate of the Variance in Sampling with Varying Probabilities},
  journal = {Journal of the Indian Society of Agricultural Statistics},
  year    = {1953},
  volume  = {5},
  pages   = {119--127}
}

@article{yates1953,
    ISSN = {00359246},
    URL = {http://www.jstor.org/stable/2983772},
    author = {Yates, F. and Grundy, P. M.},
    journal = {Journal of the Royal Statistical Society. Series B (Methodological)},
    number = {2},
    pages = {253--261},
    publisher = {[Royal Statistical Society, Oxford University Press]},
    title = {Selection Without Replacement from Within Strata with Probability Proportional to Size},
    urldate = {2026-06-15},
    volume = {15},
    year = {1953}
}

@article{liu1974,
  author  = {Liu, T. P.},
  title   = {A general unbiased estimator for the variance of a finite population},
  journal = {Sankhy\={a}: The Indian Journal of Statistics, Series C},
  year    = {1974},
  volume  = {36},
  pages   = {23--32}
}

@article{chaudhuri1978,
  title     = {On estimating the variance of a finite population},
  author    = {Chaudhuri, A.},
  journal   = {Metrika},
  volume    = {25},
  number    = {1},
  pages     = {65--76},
  year      = {1978},
  publisher = {Springer},
  doi       = {10.1007/BF02204352}
}

@article{grafstrom2013,
    author = {Grafstr{\"{o}}m, Anton and Lundstr{\"{o}}m, Niklas},
    year = {2013},
    month = {02},
    pages = {36--41},
    number = {1},
    title = {Why Well Spread Probability Samples Are Balanced},
    volume = {3},
    journal = {Open Journal of Statistics},
    doi = {10.4236/ojs.2013.31005}
}

@article{fattorini2006,
    author = {Fattorini, Lorenzo},
    year = {2006},
    month = {06},
    pages = {269--278},
    title = {Applying the Horvitz-Thompson criterion in complex designs: A computer-intensive perspective for estimating inclusion probabilities},
    volume = {93},
    number = {2},
    journal = {Biometrika},
    doi = {10.1093/biomet/93.2.269}
}

@book{wolter2007,
    author = {Wolter, Kirk M.},
    year = {2007},
    month = {01},
    edition = {2nd},
    publisher = {Springer},
    title = {Introduction to Variance Estimation},
    isbn = {978-0-387-32917-8},
    doi = {10.1007/978-0-387-35099-8}
}

@article{grafstrom2014,
    ISSN = {03036898, 14679469},
    URL = {http://www.jstor.org/stable/24586652},
    author = {Grafstr{\"{o}}m, Anton and Schelin, Lina},
    journal = {Scandinavian Journal of Statistics},
    number = {2},
    pages = {277--290},
    publisher = {[Board of the Foundation of the Scandinavian Journal of Statistics, Wiley]},
    title = {How to Select Representative Samples},
    urldate = {2026-06-15},
    volume = {41},
    year = {2014}
}

@article{grafstrom2018,
    author = {Grafstr{\"{o}}m, Anton and Matei, Alina},
    year = {2018},
    month = {04},
    pages = {792--805},
    title = {Spatially Balanced Sampling of Continuous Populations},
    volume = {45},
    number = {3},
    journal = {Scandinavian Journal of Statistics},
    doi = {10.1111/sjos.12322}
}

@misc{olofsson2023,
      title={Enhancing Precision with the Local Pivotal Method: A General Variance Reduction Approach}, 
      author={Olofsson, Marcus and Grafstr{\"{o}}m, Anton and Lundstr{\"{o}}m, Niklas L. P.},
      year={2023},
      eprint={2305.02446},
      archivePrefix={arXiv},
      primaryClass={stat.ME},
      url={https://arxiv.org/abs/2305.02446}, 
}

@article{stevens2003,
    author = {Stevens Jr, Don L. and Olsen, Anthony R.},
    year = {2003},
    month = {09},
    pages = {593--610},
    title = {Variance estimation for spatially balanced samples of environmental resources},
    volume = {14},
    number = {6},
    journal = {Environmetrics},
    doi = {10.1002/env.606}
}

@book{cochran1977,
  title     = {Sampling Techniques},
  author    = {Cochran, William G.},
  year      = {1977},
  edition   = {3rd},
  publisher = {John Wiley \& Sons},
  address   = {New York},
  isbn      = {978-0471162407}
}

@article{deville1998,
    ISSN = {00063444, 14643510},
    URL = {http://www.jstor.org/stable/2337311},
    author = {Deville, Jean-Claude and Till{\'{e}}, Yves},
    journal = {Biometrika},
    number = {1},
    pages = {89--101},
    publisher = {[Oxford University Press, Biometrika Trust]},
    title = {Unequal Probability Sampling Without Replacement Through a Splitting Method},
    urldate = {2026-06-17},
    volume = {85},
    year = {1998}
}

@article{Hartley1962,
	author = {Hartley, H. O. and Rao, J. N. K.},
	doi = {10.1214/aoms/1177704564},
	journal = {The Annals of Mathematical Statistics},
	number = {2},
	pages = {350--374},
	publisher = {Institute of Mathematical Statistics},
	title = {Sampling with Unequal Probabilities and without Replacement},
	url = {https://doi.org/10.1214/aoms/1177704564},
	volume = {33},
	year = {1962}
}

@article{Hajek1964,
	author = {Jaroslav H{\'a}jek},
	doi = {10.1214/aoms/1177700375},
	journal = {The Annals of Mathematical Statistics},
	number = {4},
	pages = {1491--1523},
	publisher = {Institute of Mathematical Statistics},
	title = {Asymptotic Theory of Rejective Sampling with Varying Probabilities from a Finite Population},
	url = {https://doi.org/10.1214/aoms/1177700375},
	volume = {35},
	year = {1964}
}

@article{tille1996,
    ISSN = {0769489X, 22726497},
    URL = {http://www.jstor.org/stable/20076043},
    author = {Till{\'{e}}, Yves},
    journal = {Annales d'{\'E}conomie et de Statistique},
    volume = {44},
    pages = {177--189},
    publisher = {[GENES, ADRES]},
    title = {Some Remarks on Unequal Probability Sampling Designs without Replacement},
    urldate = {2026-06-17},
    year = {1996}
}

@article{Brewer2003,
    title = {The high entropy variance of the Horvitz-Thompson estimator},
    author = {Brewer, Kenneth R. W. and Donadio, Martin E.},
    year = {2003},
    language = {English},
    volume = {29},
    pages = {189--196},
    journal = {Survey Methodology},
    publisher = {Statistics Canada},
    number = {2}
}

@article{mckay1979,
    author = {McKay, M. D. and Beckman, Richard J. and Conover, William J.},
    year = {1979},
    month = {05},
    pages = {239--245},
    title = {A Comparison of Three Methods for Selecting Values of Input Variables in the Analysis of Output From a Computer Code},
    volume = {21},
    number = {2},
    publisher = {[Taylor \& Francis, Ltd., American Statistical Association, American Society for Quality]},
    journal = {Technometrics},
    doi = {10.1080/00401706.1979.10489755}
}

@Manual{lhsR,
    title = {{lhs}: Latin Hypercube Samples},
    author = {Rob Carnell},
    year = {2026},
    note = {R package version 1.2.1},
    url = {https://cran.r-project.org/package=lhs},
    doi = {10.32614/CRAN.package.lhs},
  }

@MISC{simulationlib, 
 author = {Surjanovic, S. and Bingham, D.}, 
 title = {Virtual Library of Simulation Experiments: Test Functions and Datasets}, 
 howpublished = {Retrieved June 17, 2026, from \url{http://www.sfu.ca/~ssurjano}}, 
 year = {2013}
}

@book{hansen1953,
  title={Sample Survey Methods and Theory},
  author={Hansen, Morris H. and Hurwitz, William N. and Madow, William G.},
  volume={1--2},
  year={1953},
  publisher={John Wiley \& Sons},
  address={New York},
  isbn = {9780471349143}
}

@book{kish1965,
  title={Survey Sampling},
  author={Kish, Leslie},
  isbn={9780471489009},
  lccn={65019479},
  url={https://books.google.com/books?id=gp1qAAAAMAAJ},
  year={1965},
  publisher={John Wiley \& Sons},
  address = {New York}
}

@book{rubin1987,
  author = {Rubin, Donald B.},
  title = {Multiple Imputation for Nonresponse in Surveys},
  year = {1987},
  publisher = {John Wiley \& Sons},
  address = {New York},
  series = {Wiley Series in Probability and Mathematical Statistics},
  isbn = {9780471087052},
  doi = {10.1002/9780470316696},
  url = {https://doi.org/10.1002/9780470316696}
}

@article{kim2021,
    author = {Kim, Jae Kwang and Park, Seho and Chen, Yilin and Wu, Changbao},
    title = {Combining non-probability and probability survey samples through mass imputation},
    journal = {Journal of the Royal Statistical Society: Series A (Statistics in Society)},
    volume = {184},
    number = {3},
    pages = {941--963},
    doi = {10.1111/rssa.12696},
    url = {https://rss.onlinelibrary.wiley.com/doi/abs/10.1111/rssa.12696},
    eprint = {https://rss.onlinelibrary.wiley.com/doi/pdf/10.1111/rssa.12696},
    year = {2021}
}

@misc{mozer2024,
      title={More power to you: Using machine learning to augment human coding for more efficient inference in text-based randomized trials}, 
      author={Mozer, Reagan and Miratrix, Luke},
      year={2024},
      eprint={2309.13666},
      archivePrefix={arXiv},
      primaryClass={stat.ME},
      url={https://arxiv.org/abs/2309.13666}, 
}

@article{DiceKriging,
    title = {{DiceKriging}, {DiceOptim}: Two {R} Packages for the Analysis of Computer Experiments by Kriging-Based Metamodeling and Optimization},
    author = {Olivier Roustant and David Ginsbourger and Yves Deville},
    journal = {Journal of Statistical Software},
    year = {2012},
    volume = {51},
    number = {1},
    pages = {1--55},
    doi = {10.18637/jss.v051.i01},
}

@article{Frechet1935,
    author = {Fr{\'e}chet, Maurice},
    journal = {Fundamenta Mathematicae},
    language = {fre},
    number = {1},
    pages = {379--387},
    title = {G{\'e}n{\'e}ralisation du th{\'e}or{\`e}me des probabilit{\'e}s totales},
    url = {http://eudml.org/doc/212798},
    volume = {25},
    year = {1935},
}

@book{Santner2003,
  author    = {Thomas J. Santner and Brian J. Williams and William I. Notz},
  title     = {The Design and Analysis of Computer Experiments},
  publisher = {Springer},
  series    = {Springer Series in Statistics},
  year      = {2003},
  address   = {New York},
  doi       = {10.1007/978-1-4757-3799-8}
}

@book{RasmussenWilliams2006,
  author    = {Carl Edward Rasmussen and Christopher K. I. Williams},
  title     = {Gaussian Processes for Machine Learning},
  publisher = {MIT Press},
  series    = {Adaptive Computation and Machine Learning},
  year      = {2006},
  address   = {Cambridge, MA},
  isbn      = {978-0-262-18253-9}
}

\end{document}